\documentclass[aps,prc,onecolumn,superscriptaddress]{revtex4}
\usepackage[latin1]{inputenc}
\usepackage{bm}
\usepackage{epsfig}
\usepackage{amsthm}
\usepackage{amsfonts}
\usepackage{float}
\usepackage{amsmath,amssymb}
\usepackage{color}
\usepackage{slashed}
\usepackage{relsize}
\usepackage[toc,page]{appendix} 

\usepackage{graphicx}
\usepackage{dcolumn}
\usepackage{bm}
\newcommand{\be}{\begin{equation}} 
\newcommand{\ee}{\end{equation}} 
\newcommand{\bea}{\begin{eqnarray}} 
\newcommand{\eea}{\end{eqnarray}} 
\newcommand{\nn}{\nonumber} 
\newcommand{\mintedim}[2]{{\int\kern-0.50em\mbox{{\small$\mathop{\frac{\mbox{{\small${\rm d^{#2}}\vect{#1}$}}}{\mbox{{\small$(2\pi)^{#2}$}}}}$}}\ }} 
\newcommand{\inteonedim}[1]{{\int_0^\infty\kern-1em\mbox{{\small${\rm d}{#1}$}}}} 
\newcommand{\rmi}[1]{{\mbox{\scriptsize{#1}}}} 
\newcommand{\vect}[1]{\bm{#1}} 

\begin{document}
\title{
\textcolor{black} {Analytical calculations of the Quantum Tsallis thermodynamic variables}
}

\author{Ayman Hussein}
\email{s-ayman.mh@zewailcity.edu.eg}
\affiliation{Zewail City of Science, Technology and Innovation, Giza, Egypt}
%
\author{Trambak Bhattacharyya}
\email{bhattacharyya@theor.jinr.ru}
\affiliation{Bogoliubov Laboratory of Theoretical Physics, Joint Institute for Nuclear Research, Dubna, \\
Moscow Region, Russia}

\begin{abstract} 
In this article, we provide an account of analytical results related to the Tsallis thermodynamics that have 
been the subject matter of a lot of studies in the field of high-energy collisions. After reviewing the results
for the classical case in the massless limit and for arbitrarily massive classical particles, we compute the
quantum thermodynamic variables. For the first time, the analytical formula for the pressure of a Tsallis-like
gas of massive bosons has been obtained. Hence, this article serves both as a brief review of the knowledge gathered in this area, and
as an original research that forwards the existing scholarship. The results of the present paper will be important in a plethora
of studies in the field of high-energy collisions including the propagation of non-linear waves generated by the traversal of
high-energy particles inside the quark-gluon plasma medium showing the features of non-extensivity.
\\
\\
\smaller{Keywords: Tsallis statistics, Tsallis thermodynamics, thermodynamic variables, integral representation}
\end{abstract}

\maketitle
%

\section{Introduction}

Power-law distributions have been routinely used to describe particle yields in high-energy collision physics. It has been observed that 
the pions, kaons, protons (and other hadrons) originated in these collision events follow a power-law distribution in the transverse momentum ($p_{\text{T}}$) space. The power-law formula utilized by different experimental collaborations like STAR \cite{TsSTAR}, PHENIX \cite{TsPHENIX}, ALICE \cite{TsALICE} and CMS \cite{TsCMS} use the following form of a power-law transverse momentum distribution,
\bea
\frac{d^2N}{dp_{\text{T}}dy}= p_{\text{T}} \frac{dN}{dy} \frac{(n-1)(n-2)}{nC(nC+m_0(n-2))}\left(1+\frac{m_{\text{T}}-m_0}{nC}\right)^{-n},
\label{powex}
\eea
that has some correspondence with the form of the Tsallis transverse momentum distribution proposed by Cleymans and Worku in 2012 \cite{CWJPG,CWEPJA},
\bea
\frac{d^2N}{dp_{\text{T}}dy} = \frac{gV}{(2\pi)^2} p_{\text{T}} m_{\text{T}}\cosh y \left(1+(q-1)\frac{m_{\text{T}}\cosh y -\mu}{T}\right)^{-\frac{q}{q-1}}.
\label{tscl}
\eea
In Eqs.~\eqref{powex} and \eqref{tscl} $n$, $C$, $m_0$, $V$ (volume), $q$ (Tsallis parameter), $T$ (Tsallis temperature) and $\mu$ (chemical potential) are fit parameters,
$g$ is degeneracy, $m_{\text{T}}$ is transverse mass and $y$ is rapidity. Connections of these distributions can be made with the Tsallis statistical mechanics developed by C. Tsallis in 1988 \cite{Tsal88}, a statistics that has long been used to tackle a medium with fluctuation, long-range correlation \cite{WilkPRL,WilkPRC,WilkOsada,BiroMolnar}, small system size \cite{BiroSmall} and fractal structure \cite{DeppmanFractal}. It has been shown that the Tsallis-like distribution proposed in Refs.~\cite{CWJPG,CWEPJA} obey thermodynamic relations. Later on, from the definition the Tsallis entropy, the distribution in Eq.~\eqref{tscl} has been shown to be the zeroth-order approximation of the exact Tsallis-like transverse momentum distribution in the Tsallis-2 prescription \cite{ParBh19} that can be extremely useful for the LHC phenomenology \cite{CleymansPLB13,mcdprd,TripathyEPJA18,ALICE18,jpg20}. The same distribution can also be obtained from the $q$-dual statistics proposed in Ref.~\cite{Parqdual}. In the papers that spearheaded the study of Tsallis thermodynamics, there were comparisons between the Tsallis-like classical and quantum distributions and their Boltzmann-Gibbs (BG) counterpart (see, e.g. \cite{CWJPG}) that is achieved once the $q$ parameter approaches unity.  

A curious reader might also have felt an implicit necessity to be able to compare the thermodynamic variables in the two formulations -- Tsallis and BG. One might always take a numerical approach to provide an answer, as the analytical formulae of the Tsallis thermodynamic variables were not widely available, as opposed to their Boltzmann-Gibbs counterparts that are expressible in terms of the modified Bessel functions. However, this does not mean that there were no attempts to find analytical results. Lavagno in 2002 already provided these expressions in terms of the $q$-modified Bessel functions of the second kind \cite{lavagnopla}. But the properties of this group of $q$-modified special functions are not widely available and known to physicists. Hence, it was necessary to explore this question further. Thermodynamic variables are important as their relationships form the equation of state that is an important input to study, for example, evolution of the quark-gluon plasma (QGP) medium \cite{Gyulassy:1983ub,Akase:1990yd,Csernai:2003xd,Song:2008hj,Teaney:2009qa,Chaudhuri:2010in,Jaiswal:2016hex}, propagation of non-linear waves in the QGP using the hydrodynamic equation \cite { rahaplb,nlwprc,nlwprd,nlwplb,tube,bhattaepjc1,bhattaepjc2} and so on. 

There was a renewed interest in this attempt during professor Cleymans' visit to India in 2015. This attempt was based on the observation that the Tsallis-like distributions can be
written using the Taylor's series expansion in the increasing order of $(q-1)^n,~n\in\mathbb{Z}^{\geq}$ 
($\mathbb{Z}^{\geq}$ represents the set of non-negative integers). In a joint paper \cite{TsallisTaylor}, the Tsallis thermodynamic variables for an ideal gas of massive particles were explored. However, eventually it was realized that the results were restricted by the fact that the Tsallis distribution was truncated at $\mathcal{O}(q-1)^2$. Such an early truncation led to restrictions in the phase-space dictated by the ratios involving $q$, $T$ and the single particle energy $E_p$ (see also \cite{WilkOsada}). So, the question of finding an unapproximated analytical expression of the Tsallis thermodynamic variables was still open. In the meantime, one of us (T.B.) joined the group of professor Cleymans in Cape Town, and some progress ensued. Professor Cleymans proved that the calculations for the massless case can be performed analytically, and that was a breakthrough. This was the inspiration behind another paper with him \cite{bcmprd} that elaborated a method to analytically calculate the Tsallis thermodynamic variables for the massive particles without an approximation (like the Taylor's series expansion, or considering massless case and so on) using the Mellin-Barnes contour integral representation of the Tsallis distribution. It was found that the interesting features (like poles), that were missed in the Taylor's approximated calculations, are intact in the massless as well as the massive case. In this article we extend the existing knowledge to the quantum domain, and propose a method to calculate analytical formulae of the quantum Tsallis thermodynamic variables. In the present article, we have considered isotropic momentum distributions. So, the results can also be useful for the quark-gluon plasma medium formed in the early universe \cite{isodist0} or other branches of physics that use isotropic distributions \cite{isodist1,isodist2,isodist3,isodist4,isodist5}. For a more realistic scenario
to treat high-energy collision physics, anisotropy may be considered, but we reserve that for a future study.

Being involved in such a journey with professor Cleymans as a friend, philosopher and guide is truly rewarding. The present article serves as our tribute to his memory and his inspiring scientific curiosity. 


\section{Review: Tsallis thermodynamics: $m=0$, $\mu=0$}
The Tsallis thermodynamic variables can be written in terms of the Tsallis single particle distribution. The single particle distributions
can be obtained following three different averaging schemes \cite{tsallis123}, that we name Tsallis-1, 2 and 3.
These schemes differ in the definition of the mean values (for example the mean energy) utilized for the constrained maximization of the Tsallis entropy. In the
first scheme, the mean is defined as $\langle O \rangle = \sum_i p_i O_i$, in the second scheme $\langle O \rangle = \sum_i p_i^q O_i$, and in the
third scheme $\langle O \rangle = \sum_i p_i^q O_i/\sum p_i^q$, where $\{p_i\}$ are the probabilities of micro-states.
It will be worthwhile to mention that following the previous works \cite{CWJPG,CWEPJA}, we follow the second
averaging scheme. The present article focuses entirely on the analytical method to calculate the Tsallis thermodynamic variables, and based on this
prescription, it will be relatively straightforward to extend the calculations for other forms of the single particle distributions. With this understanding, 
we consider the following isotropic quantum (b: bosons; f: fermions) single particle distributions (positive sign for fermions, negative sign for bosons):
\bea
n_{\text{b}/\text{f}}= \frac{1}{\left(1+(q-1)\frac{E_p-\mu}{T}\right)^{\frac{q}{q-1}}\pm1},
\label{tsq}
\eea
where $E_p=\sqrt{p^2 +m^2}$ is the single particle energy for a particle of mass $m$ and three-momentum $p$.
The above distribution is not entirely phenomenological because it can be obtained (after certain approximations) from the constrained maximization of the Tsallis entropy as 
shown in \cite{ParBh21}. The above distribution is similar to (but not exactly the same as) the one proposed in \cite{millerTsFD,TsFDPLA}. For the sake of completeness, we also quote the popular classical (Maxwell-Boltzmann or MB) Tsallis-like single particle distribution that gives rise to Eq.~\eqref{tscl}:
\bea
n_\text{MB} =  \left(1+(q-1)\frac{E_p-\mu}{T}\right)^{-\frac{q}{q-1}}.
\eea
With the help of the single particle distributions ($n_{\text{s}};$ s=b, f, MB), the thermodynamic variables pressure ($P$), mean energy ($E$) and mean number 
of particles ($N$) can be written in the following way:
\bea
P = g \mintedim{p}{3}\frac{p^2}{3~E_\rmi{p}}\ n_{\text{s}} \label{pressure};
\ \ E = gV\mintedim{p}{3}E_\rmi{p}\ n_{\text{s}} \label{epsilon};
\ \ N = gV \mintedim{p}{3} n_{\text{s}} \label{Number}.
\eea


\subsection{Classical case}
In this subsection we tabulate the results for the classical (Tsallis Maxwell-Boltzmann) case in the massless approximation. The thermodynamic variables 
pressure ($P$), energy density ($\epsilon=E/V$) and number density ($\rho=N/V$) putting $\mu=0$ are given by \cite{bcmprd}:
\begin{align}
    P &= \frac{gT^4}{6\pi^2} \frac{1}{(2-q)(3/2-q)(4/3-q)}\\
    \epsilon &= \frac{gT^4}{2\pi^2} \frac{1}{(2-q)(3/2-q)(4/3-q)} = 3 P\\
    \rho &= \frac{gT^3}{2\pi^2} \frac{1}{(2-q)(3/2-q)}.
\end{align}
It is interesting to note from the above expressions (for example pressure) that the first pole of $q$ appears at $q=4/3$ that is close to $1.\overline{3}$ \cite{repdec}. This puts an upper-bound
on the $q$ value that is a parameter to be determined from the experimental data. Experimental observations indeed show that $q$ values do obey this upper-bound that 
is imposed because of the finite values of the thermodynamic variables. However, other considerations may further shrink the range \cite{TsQCD}. It is also 
noteworthy that the upper-bound of $q$ (say $q^{(D)}_{\text{max}}$) obtained from the thermodynamic considerations changes with the dimension of the system
as $q^{(D)}_{\text{max}}<1+1/(D-1)$. Hence, when we put $D=4$, this value is $4/3$.

\subsection{Quantum case}
Tsallis quantum thermodynamic variables in the massless limit are given by the following closed analytic formulae. The details of the calculations can be 
found in Ref.~\cite{bhattaepjc1}:

\subsubsection{Bosons}
\begin{align}
 P_\text{b} &= \frac{g T^4}{6 \pi ^2 (q-1)^3 q} \left[3 \psi
   ^{(0)}\left(\frac{3}{q}-2\right) + \psi
   ^{(0)}\left(\frac{1}{q}\right)- 3 \psi ^{(0)}\left(\frac{2}{q}-1\right) - \psi
   ^{(0)}\left(\frac{4}{q}-3\right)\right], 
   \label{Pboson} \\
   \epsilon_\text{b} &= \frac{g T^4}{2 \pi ^2 (q-1)^3 q} \left[ 3 \psi
   ^{(0)}\left(\frac{3}{q}-2\right) + \psi
   ^{(0)}\left(\frac{1}{q}\right)- 3 \psi ^{(0)}\left(\frac{2}{q}-1\right) - \psi
   ^{(0)}\left(\frac{4}{q}-3\right) \right],
   \label{epsilonboson}\\
  \rho_{\text{b}} &= \frac{g T^3}{2 \pi ^2 (q-1)^2 q} 
  \left[ 2 \psi^{(0)}\left(\frac{2}{q}-1\right) 
  - 
  \psi^{(0)}\left(\frac{3}{q}-2\right)
  - 
   \psi ^{(0)}\left(\frac{1}{q}\right)
   \right].
   \label{rhoboson}
\end{align}
\subsubsection{Fermions}
\begin{align}
    P_\text{f} =\frac{g T^4}{6 \pi ^2 (q-1)^3 q} &\Bigg[3 \Phi \left(-1,1,\frac{2}{q}-1\right)-3 \Phi
   \left(-1,1,\frac{3}{q}-2\right)+\Phi\left(-1,1,\frac{4}{q}-3\right)\nonumber\\
   &-\Phi\left(-1,1,\frac{1}{q}\right)\Bigg] ,\label{Pfermion}\\
   \epsilon_\text{f} = \frac{g T^4}{2 \pi ^2 (q-1)^3 q} &\Bigg[3 \Phi \left(-1,1,\frac{2}{q}-1\right)-3 \Phi
   \left(-1,1,\frac{3}{q}-2\right)+\Phi
   \left(-1,1,\frac{4}{q}-3\right)\nonumber\\
   &-\Phi\left(-1,1,\frac{1}{q}\right)\Bigg],
   \label{epsilonfermion} \\
   \rho_{\text{f}} = \frac{g T^3}{2 \pi ^2 (q-1)^2 q} &\left[-2 \Phi \left(-1,1,\frac{2}{q}-1\right)+ \Phi
   \left(-1,1,\frac{3}{q}-2\right)+\Phi
   \left(-1,1,\frac{1}{q}\right)\right],
\end{align}
\\
where $\psi^{(0)}(z)$ is the digamma function, and $\Phi(a,b,z)$ is the Lerch's transcendent \cite{Bateman}, both of which have poles at $z=0$. In view of this comment,
we observe that, just like the classical case, the first pole in the thermodynamic variables (for example pressure) appears at $q=4/3$. And hence, the upper-bound $q<4/3$
is still relevant. Before going to the next section, we would like to comment that the above results can be extended to treat very light particles, as discussed in Ref.~\cite{bhattaepjc1}. It is possible to obtain the $\mathcal{O}(m^2T^2)$ correction to the above approximated results that may work well for the light quarks like up and down. 

\section{Review and new results: Tsallis thermodynamics: $m\neq0$, $\mu=0$}
In this section, we quote the closed analytical formula of the pressure in a gas of massive classical and quantum particles without utilizing
any approximation. The classical case has already been considered in a previous report \cite{bcmprd}. However, we are not
aware of any other article reporting results for the quantum case with arbitrarily massive particles (albeit results are available for slightly massive particles). 
In this section, we only quote the obtained results for classical and quantum (boson) particles. Detailed mathematical steps for obtaining quantum results 
are deferred until the next section. 

\subsection{Classical case (review)}
In this subsection, we tabulate analytical results of classical Tsallis thermodynamics for arbitrarily massive particles. This calculation involves an integral representation
of a power-law function that is at the heart of the Tsallis statistics. Because of the nature of the integrals involved, the convergence conditions lead to two separate formulae 
for the thermodynamic variables for two regions $q>1+T/m$, that we call the `upper region' and $q\leq 1+T/m$, that we call the `lower region'. The origin of these regions are explained in section \ref{TsQVar} where we consider the quantum case.
\subsubsection{Upper region: $q>1+T/m$}
The analytical expression result valid for the upper region is given below:
\begin{align}
P_\rmi{U}&=\frac{g\ m^4}{16 \pi^{\frac{3}{2}}}\ \left(\frac{T}{(q-1) \ m}\right)^{\frac{q}{q-1}}\ \left[%
\frac{\Gamma\left(\frac{4-3q}{2(q-1)}\right)}{\Gamma\left(\frac{2q-1}{2(q-1)}\right)} \,%
_2F_1\left(\frac{q}{2 (q-1)},\frac{4-3q}{2 (q-1)},\frac{1}{2}; \frac{T^2}{(q-1)^2 m^2}\right)\right. \nn\\
&\ \ \ \ \ \ \ \ \ \ - \left. \frac{2T}{(q-1) m}\times\frac{\Gamma\left(\frac{3-2q}{2(q-1)}\right)}{\Gamma\left(\frac{q}{2(q-1)}\right)} \,%
_2F_1\left(\frac{2q-1}{2 (q-1)},\frac{3-2q}{2(q-1)},\frac{3}{2}; \frac{T^2}{(q-1)^2 m^2}\right)%
\right],
\label{MassiveResPU}
\end{align}
where $\Gamma(z)$ is the Gamma function, and $_2F_1(a,b,c; z)$ is the hypergeometric function \cite{Bateman}.

\subsubsection{Lower region: $q<1+T/m$}
The analytical expression result valid for the lower region is given below:
\begin{align}
P_\rmi{L}=\frac{g\ m^4}{2^{\frac{q}{q-1}}\pi^{\frac{3}{2}}} &\left[%
\frac{(q-1)^2\ (3-q)\ \Gamma\left(\frac{1}{q-1}\right)}{(4-3q)\ (3-2q)\ (2-q) \Gamma\left(\frac{1+q}{2 (q-1)}\right)}%
\right]\nn \\ 
&\ \ \ \ \ \ \ \ \ \ \ \ \ \times\,_2F_1 \left( \frac{2q-1}{2(q-1)},\frac{q}{2(q-1)},\frac{3-q}{2(q-1)}, 1- \frac{(q-1)^2~ m^2}{T^2}\right)%
.\label{MassiveResPL}
\end{align}
It is worth noticing that both the expressions in general require $q<4/3$ for the consistency of the framework apart from the (upper or lower) limits based on the convergence
criterion.


\subsection{Quantum case: bosons (new results)}
In this subsection we tabulate the newly found analytical results of the Tsallis thermodynamics for arbitrarily massive bosons. In comparison with the classical case, there is an extra
step in the quantum calculations that entails expressing the quantum single particle distributions as a superposition of an infinite number of classical distributions. 
However, all the other procedures are the same as the classical part. In the quantum case also, two analytical formulae for the upper and the lower regions have
been obtained, as given below:
\subsubsection{Upper region: $q>1+T/m$}
\begin{align}
    P_{\text{U,b}} &= \mathlarger{\mathlarger{\sum}}_{s=1}^{s_0}\frac{g m^4}{32 \pi^2 \Gamma\left(\frac{qs}{q-1}\right)}\Bigg(\frac{2T}{m(q-1)}\Bigg)^{\frac{qs}{q-1}}
    \Bigg[ \Gamma \left(\frac{q s}{2 (q-1)}\right) \Gamma \left(\frac{q s}{2 (q-1)}-2\right) \nonumber\\
   & \times \, _2F_1\left(\frac{q s}{2 (q-1)},\frac{q s}{2 (q-1)}-2;\frac{1}{2};\frac{T^2}{m^2 (q-1)^2}\right)
    - \frac{2T}{m(q-1)}
    \Gamma \left( \frac{q s -3q +3}{2(q-1)}\right) 
     \nonumber\\
    & \times \Gamma \left( \frac{q s+q-1}{2(q-1)} \right)\, _2F_1\left(\frac{q s -3q +3}{2(q-1)}, \frac{q s+q-1}{2(q-1)};\frac{3}{2};\frac{T^2}{m^2 (q-1)^2}\right)\Bigg].
    \label{Pfinal1}
\end{align}
\subsubsection{Lower region: $q<1+T/m$}
\begin{align}
    P_{\text{L,b}}  = \frac{gT^4}{16 (q-1)^4} \ &\mathlarger{\mathlarger{\sum}}_{s=1}^{s_0}
    \, _2\tilde{F}_1\left(\frac{q s}{2 (q-1)}-2, \frac{q s}{2(q-1)}-\frac{3}{2}; \frac{q s}{q-1}-\frac{3}{2};1-\frac{m^2 (q-1)^2}{T^2}\right) \nonumber\\
    &\times \sec \left(\frac{\pi  q s}{q-1}\right)\Bigg[\frac{\Gamma \left(\frac{q s-4q+4}{2 (q-1)}\right)}{\Gamma \left(\frac{q (s-5)+5}{2(1-q)}\right) \Gamma \left(\frac{q s}{2(1-q)}+\frac{1}{2}\right) \Gamma 
    \left(\frac{q s}{2(q-1)}+\frac{1}{2}\right)}\nn\\
    &-
    \frac{\Gamma \left(\frac{q s}{2(q-1)}-\frac{3}{2} \right)}{\Gamma \left(\frac{q s}{2 (q-1)}\right) \Gamma \left(\frac{q s}{2(1-q)}+1\right) \Gamma \left(\frac{q s}{2(1-q)}+3\right)}\Bigg],
    \label{Pfinal2}
\end{align}
where the regularized hypergeometric function $\, _2\tilde{F}_1\left(a,b;c;z\right)= \, _2F_1\left(a,b;c;z\right)/\Gamma(c)$. Ideally, $s_0$ is a very large number, depending on the desired agreement between the numerical and analytical results. Eqs.~\eqref{Pfinal1} and \eqref{Pfinal2} (also repeated in Eqs.~\ref{Pfinal1again} and \ref{Pfinal2again}) are the main results of the paper. 

\section{Methodology: The Pressure of a gas of bosons following the Tsallis distribution}

\label{TsQVar}

\noindent From Eq.~\eqref{pressure}, the pressure for the bosons is given by:
\bea
    P_{\text{b}} = g \mintedim{p}{3}\frac{p^2}{3~E_\rmi{p}}\ n_{\text{b}},
\eea
where $n_{\text{b}}$ is the Tsallis Bose-Einstein single particle distribution given by Eq.~\eqref{tsq}. The spherical symmetry of the integrand implies that:

\begin{equation}
    P_{\text{b}} = \frac{g}{6\pi^2}\int_{0}^{\infty}\frac{p^4}{\sqrt{m^2+p^2}}\ \frac{1}{\Big[1+(q-1)\frac{\sqrt{m^2+p^2}}{T}\Big]^\frac{q}{q-1}-1}\ dp\label{initial},
\end{equation}
where we set $\mu=0$. Now, we describe the steps to obtain Eqs.~\eqref{Pfinal1} and \eqref{Pfinal2}.


\subsection{Rescaling the integration variable}
To simplify our calculations we define $k=p/m$ so that the pressure becomes:

\begin{equation}
    P_{\text{b}}  =\frac{gm^4}{6\pi^2}\int_{0}^{\infty}\frac{k^4}{\sqrt{1+k^2}}\ \frac{1}{\Big[1+\frac{m(q-1)}{T}\sqrt{1+k^2}\Big]^\frac{q}{q-1}-1}\ dk.
\end{equation}


\subsection{Infinite summation}
Now, we observe that just like the Boltzmann-Gibbs case, the Tsallis quantum distributions can be written in terms of an infinite summation of the 
Tsallis MB distributions in the following way:
\begin{equation}
    \frac{1}{\Big[1+\frac{m(q-1)}{T}\sqrt{1+k^2}\Big]^\frac{q}{q-1}\pm1} = \mathlarger{\mathlarger{\sum}}_{s=1}^{\infty} (-1)^{a(s+1)} \Big(1+\frac{m(q-1)}{T}\sqrt{1+k^2}\Big)^{-\frac{qs}{q-1}},
\end{equation}
where $a=0$ $(a=1)$ yields the bosonic (fermionic) distribution. This step allows us to write the Tsallis pressure in a bosonic gas in a form similar to its classical counterpart, except a summation sign in front and a power index $s$ in the denominator. Hence, we obtain:
\begin{equation}
    P_{\text{b}} =\frac{gm^4}{6\pi^2} 
    \mathlarger{\mathlarger{\sum}}_{s=1}^{\infty}\int_{0}^{\infty}\frac{k^4}{\sqrt{1+k^2}}\  \frac{1}{\Big(1+\frac{m(q-1)}{T}\sqrt{1+k^2}\Big)^{\frac{qs}{q-1}}}\ dk.
    \label{Pscaled}
\end{equation}


\subsection{Contour integral representation}
Next, we use the Mellin-Barnes contour representation \cite{MB1,MB2,MB3} of the power-law function appearing in the integrand in Eq.~\eqref{Pscaled}, and follow the procedure described in Ref.~\cite{bcmprd}. A power-law function can be written as a Mellin-Barnes contour integration as follows:
\begin{equation}
    \frac{1}{(X+Y)^\lambda}=\frac{1}{2\pi i}\int_{\epsilon-i\infty}^{\epsilon+i\infty}\frac{\Gamma(-z)\Gamma(z+\lambda)}{\Gamma(\lambda)}\frac{Y^z}{X^{\lambda+z}}\ dz,
    \label{MBrep}
\end{equation}
where $\mathrm{Re}(\lambda)>0$ \& $\mathrm{Re}(\epsilon)\in (-\mathrm{Re}(\lambda),0)$ which is the case here since $\lambda= qs/(q-1)>0$ $\Leftrightarrow$ $q,s\geq1$. 
Moreover, we can take $X= m(q-1) \sqrt{1+k^2}/T$ and $Y=1$, or other combinations, all of which would yield the following expression:
\begin{align}
    P_{\text{b}} = \frac{gm^4}{12\pi^3 i}
    \mathlarger{\mathlarger{\sum}}_{s=1}^{\infty}\Bigg(\frac{T}{m(q-1)}\Bigg)^{\frac{qs}{q-1}}
    &\int_{\epsilon-i\infty}^{\epsilon+i\infty}\frac{\Gamma(-z)\ \Gamma\left(z+\frac{qs}{q-1}\right)}{\Gamma\left(\frac{qs}{q-1}\right)}\Bigg(\frac{T}{m(q-1)}\Bigg)^{z}\ dz\nn\\
    &\times\int_{0}^{\infty}k^4(1+k^2)^{-\frac{z}{2}-\frac{1}{2}-\frac{qs}{2(q-1)}}\ dk.
\end{align}


After performing the $k$-integration, the pressure can now be written as:
\begin{align}
    P_{\text{b}} = \frac{gm^4}{32\pi^{\frac{5}{2}} i}
    \mathlarger{\mathlarger{\sum}}_{s=1}^{\infty}\Bigg(\frac{T}{m(q-1)}\Bigg)^{\frac{qs}{q-1}}
    &\int_{\epsilon-i\infty}^{\epsilon+i\infty}\frac{\Gamma(-z) 
    \Gamma\left(z+\frac{qs}{q-1}\right)
    \Gamma\left(\frac{qs}{2(q-1)}+\frac{z}{2}-2\right)}
    {\Gamma\left(\frac{qs}{q-1}\right) 
    \Gamma\left({\frac{qs}{2(q-1)}+\frac{z}{2}+\frac{1}{2}}\right)}\nn\\
    &\times\Bigg(\frac{T}{m(q-1)}\Bigg)^{z}\ dz.
\end{align}
The convergence of the scaled momentum integration requires $\mathrm{Re}(z)\geq0$.


\subsection{Wrapping contour clockwise: $q>1+\frac{T}{m}$}

To clearly identify the poles in order to get the residues of the integrand, we will send $z$ to $2z$, use the Legendre's duplication formula \cite{wol1} and Cauchy's residue
formula \cite{arfken} so that the pressure now becomes:

\begin{align}
    P_{\text{b}} =& \frac{gm^4}{64\pi^{\frac{7}{2}} i}
    \mathlarger{\mathlarger{\sum}}_{s=1}^{\infty}
    \frac{2^{\frac{qs}{q-1}}}{\Gamma \left(\frac{qs}{q-1}\right)}
    \Bigg(\frac{T}{m(q-1)}\Bigg)^{\frac{qs}{q-1}}
    \int_{\epsilon-i\infty}^{\epsilon+i\infty}
    \Gamma(-z) \Gamma\left(-z+\frac{1}{2}\right) \nn\\
     & \times \Gamma\left(z+\frac{qs}{2(q-1)}\right)
    \Gamma \left(\frac{qs}{2(q-1)}+z-2\right) \Bigg(\frac{T}{m(q-1)}\Bigg)^{2z}
    dz
    \nn\\
    =& (-2\pi i) \times \frac{gm^4}{64\pi^{\frac{7}{2}} i} \mathlarger{\mathlarger{\sum}}_{s=1}^{\infty}
    \frac{2^{\frac{qs}{q-1}}}{\Gamma \left(\frac{qs}{q-1}\right)}
    \Bigg(\frac{T}{m(q-1)}\Bigg)^{\frac{qs}{q-1}}\nn\\
    &\times 
    \mathlarger{\sum}_{\ell=0}^{\infty} 
   \left\{
   \mathrm{Res}^{(1)}\left[f(z),z=\ell\right]+ \mathrm{Res}^{(2)}\left[f(z), z=\ell+\frac{1}{2}\right]
   \right\},
   \label{Pres}
\end{align}
where $f(z)$ is defined as:
\begin{equation*}
    f(z)\equiv\Gamma (-z) \Gamma \left(\frac{1}{2}-z\right) \Gamma \left(\frac{q s}{2 (q-1)}+z\right) \Gamma \left(\frac{q s}{2 (q-1)}+z-2\right) \left(\frac{T}{m(q-1)}\right)^{2 z},
\end{equation*}
and we wrap the contour clockwise so that residues get contribution only from the poles of $\Gamma(-z)$ at the positive integers (Res$^{(1)}$) including zero, and the poles of $\Gamma(-z+\frac{1}{2})$ at the positive half-integers (Res$^{(2)}$). This clockwise wrapping of contour imposes the convergence condition $q>1+T/m$ when $z\rightarrow \infty$.
Res$^{(1)}$ and Res$^{(2)}$ are given below:
\begin{align}
   \mathrm{Res}^{(1)} =& \mathrm{Res}\big[f(z),
    \{z=\ell \ \ni \ \ell \in \mathbb{Z}^{\geq} \}\big] \nn\\
    =&
    \frac{(-1)^{\ell +1} \left(\frac{T}{m(q-1)}\right)^{2 \ell }}{\ell!}
    \Gamma \left(\frac{1}{2}-\ell \right)\Gamma \left(\ell +\frac{q s}{2 (q-1)}-2\right) \Gamma \left(\ell +\frac{q s}{2 (q-1)}\right)
    \label{res1}
    \\
    \mathrm{Res}^{(2)} =&
     \mathrm{Res} \big[f(z),
    \{z=\ell +\frac{1}{2} \ni \ \ell \in \mathbb{Z}^{\geq} \}\big] \nn\\
    =&
    \frac{(-1)^{\ell +1} \left(\frac{T}{m(q-1)}\right)^{2 \ell +1}}{\ell !}
    \Gamma \left(-\ell -\frac{1}{2}\right) \Gamma \left(\ell +\frac{q s}{2 (q-1)}-\frac{3}{2}\right) \Gamma \left(\ell +\frac{q s}{2 (q-1)}+\frac{1}{2}\right)
    \label{res2}
\end{align}

%

When we put Eqs.~\eqref{res1} and \eqref{res2} in Eq.~\eqref{Pres}, the infinite summation over $\ell$ can be expressed in terms of the hypergeometric function
$\,_2F_1$ \cite{Bateman}, and we obtain the pressure in the region $q>1+T/m$ given by Eq.~\eqref{Pfinal1}:
\begin{align}
    P_{\text{U,b}} &= \mathlarger{\mathlarger{\sum}}_{s=1}^{s_0}\frac{g m^4}{32 \pi^2 \Gamma\left(\frac{qs}{q-1}\right)}\Bigg(\frac{2T}{m(q-1)}\Bigg)^{\frac{qs}{q-1}}
    \Bigg[ \Gamma \left(\frac{q s}{2 (q-1)}\right) \Gamma \left(\frac{q s}{2 (q-1)}-2\right) \nonumber\\
   & \times \, _2F_1\left(\frac{q s}{2 (q-1)},\frac{q s}{2 (q-1)}-2;\frac{1}{2};\frac{T^2}{m^2 (q-1)^2}\right)
    - \frac{2T}{m(q-1)}
    \Gamma \left( \frac{q s -3q +3}{2(q-1)}\right) 
     \nonumber\\
    & \times \Gamma \left( \frac{q s+q-1}{2(q-1)} \right)\, _2F_1\left(\frac{q s -3q +3}{2(q-1)}, \frac{q s+q-1}{2(q-1)};\frac{3}{2};\frac{T^2}{m^2 (q-1)^2}\right)\Bigg].
    \label{Pfinal1again}
\end{align}
when we truncate the infinite summation at $s=s_0$. 


\subsection{Analytic Continuation: $q<1+\frac{T}{m}$}
In stead of keeping the dimension of the momentum space arbitrary and analytically continuing the integrand prior to wrapping (since it does not lead to a closed form),
we analytically continue the result obtained in Eq.~\eqref{Pfinal1again} using Ref. \cite{Bateman} and obtain the following result in the complementary (lower) region:
%
\begin{align}
    P_{\text{L,b}}  = \frac{gT^4}{16 (q-1)^4} \ &\mathlarger{\mathlarger{\sum}}_{s=1}^{s_0}
    \, _2\tilde{F}_1\left(\frac{q s}{2 (q-1)}-2, \frac{q s}{2(q-1)}-\frac{3}{2}; \frac{q s}{q-1}-\frac{3}{2};1-\frac{m^2 (q-1)^2}{T^2}\right) \nonumber\\
    &\times \sec \left(\frac{\pi  q s}{q-1}\right)\Bigg[\frac{\Gamma \left(\frac{q s-4q+4}{2 (q-1)}\right)}{\Gamma \left(\frac{q (s-5)+5}{2(1-q)}\right) \Gamma \left(\frac{q s}{2(1-q)}+\frac{1}{2}\right) \Gamma 
    \left(\frac{q s}{2(q-1)}+\frac{1}{2}\right)}\nn\\
    &-
    \frac{\Gamma \left(\frac{q s}{2(q-1)}-\frac{3}{2} \right)}{\Gamma \left(\frac{q s}{2 (q-1)}\right) \Gamma \left(\frac{q s}{2(1-q)}+1\right) \Gamma \left(\frac{q s}{2(1-q)}+3\right)}\Bigg],
    \label{Pfinal2again}
\end{align}
where the definition of the regularized hypergeometric function $\, _2\tilde{F}_1(a,b;c;z)$ is provided below Eq.~\eqref{Pfinal2}.

\section{Results and Discussion}

Now, some comments about the comparison of numerical results with those obtained from the analytical formulae are in order. Let us check the massless limit first.
We notice that the final result works pretty well for the case of massless particles ($m=0$), as it should. To check that this works, we substitute numerical values in Eq.~\eqref{Pfinal2}, and compare with the numerical value of the integral in Eq.~\eqref{initial}. We take, for example, $(q = 1.2,\ g = 1,\ T = 0.08~\text{GeV},\ m = 10^{-7}~\text{GeV})$ such that $q=1.2<<1+T/m$. This condition implies that we use Eq.~\eqref{Pfinal2again}. When we put $s_0=20$, both the numerical and the analytical results (obtained also from Eq.~\ref{Pboson}) agree up to eleven significant digits and the value of pressure stands to be 2.20098$\times10^{-5}$ GeV$^4$.

Next, let us consider light particles like the positively-charged pions (mass 0.140 GeV) produced at the LHC in p-p collisions. We observe that for $q=1.154$, $g=1$ and $T=0.0682$ GeV (values taken from \cite{CWJPG}), the value of $s_0$ significantly differs from the massless case when we consider the pions. For the pions ($q=1.154<1+T/m=1.487$), a similar agreement between the analytical and numerical results can be reached for $s_0=5$ and the pressure turns out to be 5.4318$\times10^{-6}$ GeV$^4$.

We also consider more massive particles like the protons (mass 0.938 GeV) produced at the LHC in p-p collisions. For $q=1.107$, $g=2$ and $T=0.073$ GeV (values taken from \cite{CWJPG}), a similar agreement between the numerical and analytical results, both of which stand to be 3.5597$\times10^{-7}$ GeV$^4$, can be obtained including just two terms, $\it{i.e.}$ $s_0=2$. It is noteworthy that in this case we use Eq.~\eqref{Pfinal1again} (with a
proper pre-factor), as $q=1.107>1+T/m\approx1.078$. In these examples, the heavier the particle, the faster the infinite summation converges. This trend is repeated when we change only mass, keeping $q$ and $T$ values unaltered.

For a gas of positively charged pions produced at the RHIC ($\sqrt{s}_{\text{NN}}=200$ GeV Au-Au, $q=1.090$ and $T=0.117$ GeV \cite{rajendra}), the pressure is 3.0089$\times 10^{-5}$ GeV$^4$. Considering all the charged particles produced at the LHC ($\sqrt{s}_{\text{NN}}=2760$ GeV Pb-Pb, $q=1.135$ and $T=0.096$ GeV \cite{jpg20}), the pressure is 6.5733$\times 10^{-5}$ GeV$^4$.  

We conclude from the comparison of numerical and analytical results that the latter works considerably well. We hope that the main results reported in this paper will sufficiently reduce the overall computation time. We have checked that for some of the above examples, computation time is almost ten times reduced when the analytical formulae are used. 

\section{Summary, conclusions, and outlook}

In summary, we have provided a brief review of the studies related to the Tsallis thermodynamics that may be important in the studies of the quark-gluon plasma and many other
systems that display fluctuation and long-range correlation, and we have presented a detailed description of how to extend those existing findings to the quantum domain (Eqs.~\ref{Pfinal1again} and \ref{Pfinal2again}). We have
used the contour integral representation of the power-law function and followed the ritual proposed in \cite{bcmprd}, after expressing the quantum distributions in terms of an infinite summation of classical (MB) distributions. We have elaborated the analytical computation of the pressure of a bosonic gas following the Tsallis statistics, and the final result can be expressed as a summation that appears from the superposition of classical distributions. However, we have noticed that in the examples discussed, only a finite number of terms are needed, and the number of required terms for convergence decreases with mass (when $q$ and $T$ are kept unaltered). The integral representation, also known as the Mellin-Barnes representation, has extensively
been used in the studies involving loop calculations in quantum field theory \cite{MB3}. Hence, in a way, this is one of the examples where techniques established in one field of research 
benefit another. Although not mentioned in the article, extension to the fermionic case is straightforward. The only difference in summation comes owing to a factor $(-1)^{s+1}$ appearing with each term. In this paper, we have provided the results only of the pressure of a Tsallis-like bosonic gas. Other thermodynamic variables of such
a system can be calculated by appropriately differentiating pressure. Extension to the $\mu\neq0$ case can be performed with a proper identification of $X$ and $Y$ in Eq.~\eqref{MBrep}. Also, in this case, the convergence condition for clockwise wrapping is modified \cite{bcmprd}.

There may be numerous applications of the present work, but we would like to mention a particular field that has caught some recent interest. Of late, there have been studies \cite{bhattaepjc1,bhattaepjc2} reporting the propagation of non-linear waves in the quark-gluon plasma fluid (both ideal and viscous) in which constituents follow the Tsallis-like distributions. In those articles, a Tsallis-like MIT bag equation of state, considering massless (or very light) particles, has been used. It will be interesting to modify the equation of state incorporating the present findings. It will also be interesting to extend the study for hadronic gases. It has been shown \cite{ParBh19} that the exact Tsallis single particle distribution is expressed in terms 
of a series summation, and the distributions used in Eq.~\eqref{tsq} are only the approximations. For low-energy collisions (e.g. in the NICA and FAIR experiments), terms beyond the one used in the present work may be important. It will be worthwhile to investigate how those additional terms would affect the present results, and hence, the studies utilizing them.

\vspace{6pt}

\section*{Acknowledgements}
A. H. acknowledges all-round support from Alia Dawood during this work and 
stimulating discussions with Mohamed Elekhtiar and Mohamed Al Begaowe. T. B. acknowledges partial support from the joint project
between the JINR and IFIN-HH.  T. B. also gratefully acknowledges discussions with 
Sylvain Mogliacci regarding the intricacies of the Mellin-Barnes representation used in the paper as well as generous 
support from the University of Cape Town where the foundation of this work was prepared. Authors thank Rajendra Nath Patra
for providing fit parameter values of RHIC data.

\end{document}